# From Service-Oriented Computing to Metaverse Services: A Framework for Inclusive and Immersive Learning for Neurodivergent Students

Rachid Hamadi[1], Abdelmounaam Rezgui[2], and Ali Darejeh[1]

[1]School of Computer Science and Engineering, UNSW, Sydney, Australia
`r.hamadi@unsw.edu.au, ali.darejeh@unsw.edu.au`
[2]School of Information Technology, Illinois State University, Normal, IL 61790 USA
`arezgui@ilstu.edu`

**Abstract.** The metaverse offers immersive and adaptive learning environments for neurodivergent students to thrive and reach their full potential. In this paper, we propose a generic framework that leverages metaverse services as an evolution beyond traditional service-oriented computing, enabling more interactive, personalized, and engaging educational experiences. By integrating AI-driven adaptability, multimodal interaction, and privacy-first service design, the framework ensures that learning remains accessible, inclusive, and secure. Additionally, we explore the challenges associated with scalability, data privacy, and ethical considerations while highlighting opportunities for fostering safe and student-centered virtual spaces. Our analysis underscores the potential of metaverse-based learning to bridge accessibility gaps, support social-emotional development, and empower neurodivergent learners in both digital and real-world settings. We also provide recommendations and policy considerations for creating a secure, inclusive, and scalable metaverse learning ecosystem for neurodivergent students.

## 1 Introduction

Neurodivergence is globally prevalent. Estimates suggest that 15%-20% of the world's population have some form of neurodivergence [1]. A sizable proportion of students worldwide are neurodivergent. For example, in Australia, 1 in 10 school students have a disability, many of whom have neurological disability [2]. In higher education, a report by the Australian Centre for Student Equity and Success (ACSES) shows that, in 2023, 12.7% of undergraduate students reported having a disability, with "Mental health condition", being listed by approximately 49.5% of students with disability [3]. In the US, a CDC (Centers for Disease Control and Prevention) report indicates that approximately 11.4% of US children aged 3-17 have been diagnosed with Attention-Deficit/Hyperactivity Disorder (ADHD) [4]. According to the Yale

Center for Dyslexia and Creativity, Dyslexia, which is the most common neurocognitive disorder, affects 20% of the US population and represents 80%-90% of all those with learning disabilities [5].

Neurodivergent students, including those with autism, ADHD, and dyslexia, often struggle with traditional educational environments due to rigid structures, sensory overload, and a lack of personalized learning methods [6]. These environments can often be overwhelming and challenging for neurodivergent students. For two decades, service-oriented computing (SOC) has facilitated the modular and flexible development of digital education tools, improving accessibility and efficiency [7]. However, SOC-based learning platforms often face challenges in delivering inclusive and immersive experiences needed for neurodivergent students to succeed [8]. The emergence of metaverse technologies is a promising development that can enable the development of tailored, immersive, and flexible learning platforms that cater to the individual needs of neurodivergent students. We believe that the shift to metaverse services [9] is the next step in the evolution of educational technology, offering real-time virtual ecosystem learning environments encompassing AI-driven personalization that can be tailored to individual cognitive and sensory needs [10]. These services integrate the latest advancements in virtual reality (VR), augmented reality (AR), artificial intelligence (AI), biosensing, and cloud computing, making education more interactive and inclusive [11]. Additionally, the evolution of the Internet towards Web 3.0, also known as the semantic web, characterized by decentralization, immersive learning, global accessibility, and personalized learning experiences, offers new opportunities for inclusive education.

Conventional SOC online platforms cannot match immersive, adaptive, and persistent digital learning environments that metaverse services offer [12]. VR and AR technologies enable embodied and multi-sensory interactions that enhance comprehension and reduce cognitive overload for neurodivergent learners who may struggle with abstract materials and rigid structures [13]. These learning environments create a powerful sense of presence, where learners feel physically set within the virtual space when experienced through VR headsets, thus increasing focus, emotional connection to the content, and overall engagement. Features like haptic feedback, spatial audio, and adjustable sensory settings make learning more accessible, comfortable, and immersive [14]. By leveraging real-time and AI-driven personalization, these learning environments dynamically adapt content and feedback to each student's needs, hence supporting diverse cognitive profiles and emotional states [15].

Beyond individualized learning, the metaverse facilitates meaningful collaboration through shared virtual spaces such as classrooms and laboratories. These environments support real-time teamwork on projects, simulations, and design tasks, regardless of geographic location [16]. Tools like avatar-based communication, interactive whiteboards, and co-creation platforms allow students to collaborate while managing their social interaction preferences, that is, an important consideration for learners with sensory sensitivities or social anxiety [17]. The sense of co-presence in a shared virtual space not only enhances engagement but also nurtures social skills, fostering inclusion and a sense of community among peers.

Crucially, metaverse learning environments are persistent and context-aware, meaning they evolve with the learner over time [18]. Unlike static educational systems, they retain user preferences and learning histories, enabling continuous, scaffolded progression. Integration with Web 3.0 technologies further empowers students through decentralized, secure, and user-owned data management, hence supporting consistent accommodation across platforms [19]. Together, these capabilities create a flexible, immersive, and inclusive learning ecosystem ideally suited to the needs of neurodivergent students.

Recent studies show that the metaverse is showing exciting possibilities for immersive and inclusive education. Some have explored how virtual reality and layered classroom models can make learning more engaging and adaptable, especially for neurodivergent students [20], [21]. Others highlight the importance of designing virtual spaces that support social inclusion and accessibility [22], [23], [24]. These works help build the case for using metaverse technologies to create more personalized and student-centered learning environments. However, while these studies offer valuable insights, many fall short when it comes to privacy, security, and ethical service design. That is where our framework steps in by putting privacy-first design and AI-driven adaptability at the core, we aim to create metaverse services that are not only inclusive and immersive but also secure and ethically sound.

In this paper, we present a framework for inclusive and immersive learning for neurodivergent students. We argue that transitioning from SOC services to metaverse services within the Web 3.0 framework can significantly enhance learning experiences for neurodivergent students by providing adaptive, immersive, and secure learning environments. Emerging real-time metaverse technologies will enable learning systems offering near-instantaneous synchronization between physical and virtual worlds [1]. We also elaborate on the key challenges in the path towards the vision of the proposed framework. These challenges include service scalability and interoperability, data privacy and security, digital equity and accessibility, cognitive load management, and infrastructure demands [2].

The rest of the paper is organized as follows. Section 2 discusses transitioning from SOC to metaverse services. Section 3 presents the generic framework for metaverse-based neurodivergent learning. Section 4 describes opportunities of metaverse learning for neurodivergent students. Section 5 discusses challenges and risks. Section 6 provides recommendations and policy considerations. Finally, Section 6 concludes the paper and discusses future directions.

## 2    From Service-Oriented Computing to Metaverse Services

The evolution of digital learning environments has been shaped by advancements in service-oriented computing (SOC), which provides modular, reusable, and scalable educational technologies. These systems have enabled institutions to deploy cloud-based learning management systems (LMSs), adaptive tutoring platforms, and collab-orative tools that support remote and personalized learning. However, as educational needs become more immersive and interactive, particularly for neurodivergent students, SOC faces limitations in delivering deeply engaging and adaptive learning experiences. The emergence of metaverse services presents a transformative shift, integrating virtual and augmented reality with AI-driven personalization to create dynamic, interactive, and student-centered learning spaces. This section explores the role of service-oriented computing in education and how metaverse services represent the next evolutionary step in delivering immersive, adaptive, and accessible learning experiences.

## 2.1   Service-Oriented Computing in Education

In the context of SOC, a (web) service is a self-contained, reusable unit of software functionality that provides specific tasks or data to clients over a network, using standardized communication protocols.

SOC has been instrumental in modernizing digital education by enabling cloud-based platforms, reusable service components, and scalable solutions [25]. However, its limitations include:

- Lack of immersive and interactive features [26][27].
- One-size-fits-all learning approaches that do not accommodate diverse cognitive needs [28].
- Limited real-time adaptability to student responses and learning progress [29].

## 2.2   Metaverse Services as the Next Evolutionary Step

Metaverse services build upon SOC principles while offering:

- **Immersive Learning Environments:** Virtualized classrooms and adaptive content that provide students with engaging and distraction-free spaces [30].
- **AI-Driven Personalization:** Adjustments based on real-time student responses and learning analytics [31].
- **Multi-Sensory Engagement:** Integration of VR/AR, haptic feedback, and spatial audio for richer educational experiences [32].
- **Real-Time Adaptation:** Context-aware educational experiences that dynamically change based on student needs [33].
- **Cross-Platform Integration:** Seamless transitions across multiple devices and services while ensuring accessibility [34].
- **Enhanced Service-Oriented Capabilities:** The ability to dynamically configure services in immersive environments for personalized learning [35].

- **Collaborative** Working Environment: Enabling a group of users to work together within a simulated environment, interacting with one another through their own avatars [36].

Table 1 shows the mapping of SOC features to metaverse services.

**Table 1.** SOC Features to Metaverse Mapping

| SOC Feature | Metaverse Service Evolution | Impact on Neurodivergent Learning |
|---|---|---|
| Modular, Reusable Services | Virtualized, Adaptive Environments | Personalized Learning Spaces |
| Cloud-Based Infrastructure | Edge + AI-Driven Adaptation | Real-Time Adjustments to Cognitive Needs |
| API-Driven Integration | Cross-Platform Immersive Services | Seamless Access to Learning Tools |
| Stateless Transactions | Persistent, Context-Aware Sessions | Continuity in Learning Progress |
| UI-Based Interactions | Embodied, Multi-Sensory Learning | Enhanced Engagement and Understanding |

Metaverse services pose significant privacy threats since they retain sensitive user data, thus increasing the risk of unauthorized data exposure. A key challenge is designing new privacy preservation methods that ensure data privacy during metaverse services training and deployment processes without sacrificing the above offerings and efficiency.

## 3 A Generic Framework for Metaverse-Based Neurodivergent Learning

The generic framework we are proposing aligns with the core principles of SOC while leveraging metaverse technologies to create personalized, adaptive, and immersive learning experiences for neurodivergent students. This framework supports AI-driven personalization, real-time adaptability, multi-sensory engagement, and secure infrastructure to ensure inclusivity and accessibility.

The framework consists of five key components which are described below.

### 3.1 Personalized Learning Environments

One of the most critical elements of neurodivergent learning is personalization, as traditional education often fails to accommodate different cognitive processing styles.

The metaverse enhances personalization by utilizing:

- **AI-Driven Content Adaptation:** Learning pathways dynamically adjust based on the student's interactions, responses, and engagement levels. AI algorithms continuously analyze student performance to refine the curriculum, ensuring that concepts are neither too challenging nor too simplistic.

- **Customizable Sensory Settings:** The learning environment can be adjusted based on sensory sensitivities, for instance, reducing visual clutter for autistic learners, modifying sound levels, or changing color contrasts for dyslexic students.

- **Pacing Flexibility:** Unlike rigid course schedules, students can progress at their own speed, hence reducing anxiety and cognitive overload.

By implementing these features related to personalized learning environments, the metaverse promotes inclusive and adaptive educational experience, thus empowering neurodivergent learners to study in a way that suits them best.

From an implementation perspective, such personalization can be achieved through reinforcement learning algorithms (e.g., Q-learning and Deep Q-Networks) that dynamically adjust task difficulty, or collaborative filtering techniques to recommend relevant resources. Additionally, transformer-based NLP models can be integrated into conversational tutors to provide adaptive, real-time support aligned with student preferences.

### 3.2 Multimodal Interaction and Quality of Experience

A key advantage of the metaverse over traditional digital learning platforms is the ability to engage students through multimodal interaction. This includes:

- **VR/AR-Based Learning Simulations**: Students interact with three-dimensional (3D) models, virtual labs, or historical reenactments, hence improving conceptual understanding through experiential learning.

- **Haptic Feedback and Gesture-Based Interaction**: For students who struggle with text-heavy materials, haptic and motion-based interfaces provide an alternative way to interact with content, consequently improving engagement and retention.

- **Adaptive Feedback Mechanisms**: Real-time feedback from AI tutors, chatbots, or virtual mentors can provide instant guidance when students face difficulties. This reduces frustration and enhances motivation.

By leveraging the above multimodal learning approaches, education metaverse services caters to diverse learning preferences, ensuring a high Quality of Experience (QoE) for neurodivergent students.

In practice, multimodal interaction can be supported by AI-driven gesture recognition (e.g., convolutional neural networks for motion tracking), haptic feedback algorithms embedded in VR APIs, and adaptive speech recognition models that adjust sensitivity to diverse vocal patterns.

### 3.3　Service Modeling and Delivery Mechanisms

This component ensures that educational content and interactions are delivered efficiently and adaptively through a combination of cloud-based and edge computing technologies.

- **Cloud-Based AI Models**: AI-powered virtual tutors, learning analytics, and adaptive assessment tools operate seamlessly in the background, hence allowing real-time personalization.
- **Edge Computing for Low-Latency Interactions**: Processing student input locally reduces lag in VR/AR-based learning, ensuring a smooth, responsive experience.
- **Gamified Learning Elements**: Incorporating game-based learning elements, for instance, interactive quests, rewards, and point systems, will enhance motivation and will encourage active participation.
- **Dynamic Reconfiguration of Learning Services**: Unlike static learning modules, the metaverse can dynamically adapt content based on a student's real-time performance, interests, and cognitive preferences.

By dynamically personalizing the educational experience to meet neurodivergent students' unique needs, this adaptive service-oriented approach enhances engagement and improves learning outcomes.

Technically, such adaptability can be realized through edge-cloud orchestration platforms (e.g., Kubernetes with edge extensions) to minimize latency in VR/AR delivery. Game engines such as Unity or Unreal can integrate reinforcement learning APIs to reconfigure scenarios in real time, while AI-driven assessment models continuously refine learning paths.

### 3.4　Privacy-Preserving and Secure Infrastructure

Ensuring the safety and security of student data is crucial, especially when using AI-driven learning environments that collect behavioral and cognitive data. The framework integrates:

- **Federated Learning Techniques**: AI models are trained locally on student devices without sharing raw data, ensuring privacy protection while still allowing personalized learning.
- **Secure Identity Management**: Blockchain-based identity verification can protect student credentials while ensuring safe access to metaverse learning platforms.

- **Compliance with Data Protection Policies**: The framework aligns with major privacy regulations, such as GDPR (General Data Protection Regulation) and FERPA (Family Educational Rights and Privacy Act) to safeguard student sensitive information.

- **Parental and Educator Controls**: Guardians and teachers can monitor student engagement and adjust accessibility settings to ensure a safe and thriving learning environment.

These security measures balance personalization with data privacy, thus ensuring trust and transparency in metaverse-based education for neurodivergent students.

This can be implemented using federated learning protocols such as FedAvg or FedProx, where model updates, rather than raw data, are shared. To further enhance privacy protection, differential privacy techniques, such as Laplace or Gaussian noise injection, can be applied to the aggregated outputs, making it harder to identify individual data contributors. For secure identity management, blockchain-based smart contracts offer a decentralized solution for authentication and credential verification, reducing reliance on centralized systems and enhancing trust.

### 3.5 Cross-Domain Interoperability and Integration

For the metaverse to be a scalable and widely accessible learning platform, it must seamlessly integrate with existing educational systems and accommodate diverse accessibility needs. The main elements of this last component are:

- **API-Driven Compatibility with Common LMSs**: Metaverse learning platforms can connect with traditional LMS tools, such as Moodle, Canvas, or Blackboard, to provide a hybrid learning experience.

- **Standardized Accessibility Frameworks**: Ensuring that metaverse environments comply with universal accessibility standards, such as Web Content Accessibility Guidelines (WCAG) 2.2 [37], that allow students with disabilities to interact effectively.

- **Cross-Device Synchronization**: Learning progress should persist across multiple devices, enabling students to access content from VR headsets, tablets, computers, or smartphones.

- **Collaborative Learning Spaces**: Integration with social learning features, such as study groups, peer tutoring, and discussion forums, enhance community-based learning while maintaining accessibility.

By ensuring interoperability, the framework facilitates scalability, inclusivity, and adaptability, allowing a seamless transition between digital and immersive learning experiences.

Implementation can leverage open interoperability standards and APIs, such as Learning Tools Interoperability (LTI), to connect with existing LMSs. Middleware protocols can synchronize progress data across devices, while blockchain-backed credentialing ensures secure transferability of achievements between institutions.

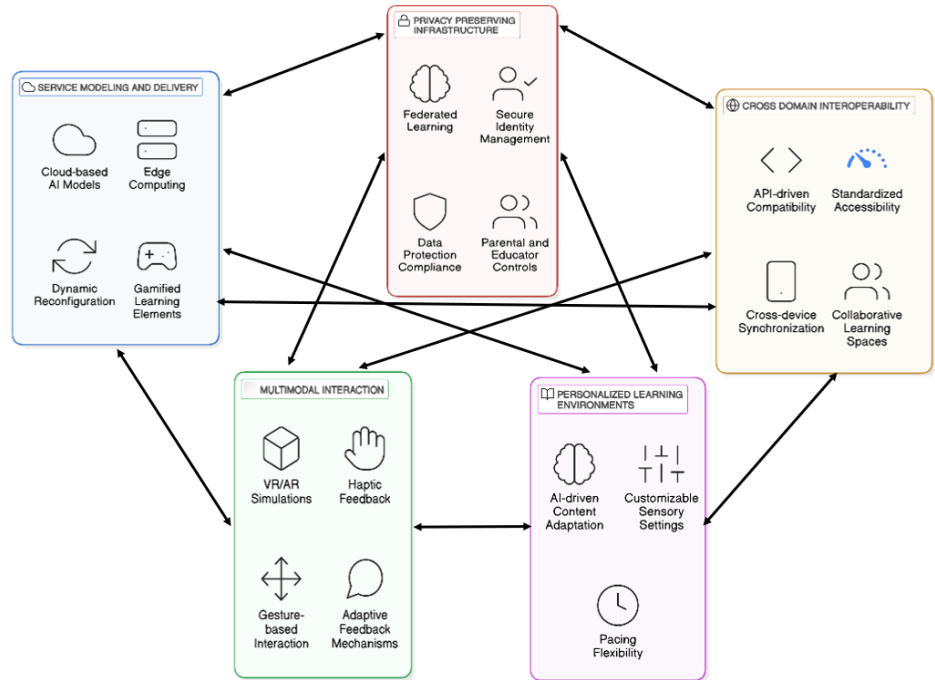

**Fig. 1.** Generic framework for metaverse-based neurodivergent learning.

Note that all five components of the proposed framework are interconnected, as shown in Fig. 1, creating a cohesive ecosystem that enhances immersive and interactive learning experiences for neurodivergent students.

The *Personalized Learning Environments* component relies on *Multimodal Interaction and QoE* component to ensure that students receive content in formats that align with their sensory and cognitive preferences, whether through visual, auditory, or haptic feedback. This dynamic interaction is supported by *Service Modeling and Delivery Mechanisms* component, which leverage cloud and AI-driven technologies to adapt content delivery in real time based on student progress and engagement levels. To maintain the security and privacy of these personalized experiences, *Privacy-Preserving and Secure Infrastructure* component plays a crucial role by implementing federated learning, decentralized authentication, and encryption protocols to protect sensitive student data. Lastly, *Cross-Domain Interoperability and Integration* component ensures that the metaverse services seamlessly interact with existing educational platforms, assistive technologies, and institutional systems, preventing data silos and ensuring a unified learning experience.

The synergy between these five components fosters an adaptive, secure, and scalable educational ecosystem where neurodivergent students can thrive. Without secure infrastructure, personalized learning environments could pose privacy risks. Without multimodal interaction, students might struggle to engage effectively with content. Moreover, without interoperability, educational institutions would face challenges in integrating metaverse services with their current systems. Thus, the proposed framework functions as an integrated model where each component reinforces the effectiveness of the others, ultimately improving learning outcomes and inclusivity.

## 4 Metaverse Learning Opportunities for Neurodivergent Students

For neurodivergent students, a metaverse learning environment conducive to proper learning must enable the following requirements.

### 4.1 Adaptive and Flexible Learning

Neurodivergent students have a wide spectrum of specific needs. A metaverse learning environment can be customized to specific needs.

**AI-Driven Customization of Lesson Plans.**

In the US, schools use Individual Education Plans (IEPs) to educate neurodivergent students. IEPs are valuable and do help special education students with their learning. IEPs also have several limitations. For example, schools may not have an adequate budget to develop IEPs. Developing an IEP is often a lengthy process that may take several weeks or months. Also, IEPs are often long and difficult to understand, and to be effective, they require proper collaboration amongst several people including (but not limited to) parents, special education teachers, psychologists, and therapists. Moreover, IEPs often set low expectations and tend to oversimplify material which results in the students with disabilities not receiving a learning comparable to that provided to students without disabilities.

A key factor that often renders IEP-based learning ineffective is that they are written based on the student's long-term disability and do not adjust, in real time, to the student's current learning ability. Research shows that a student's learning ability decreases under both psychological and physiological stress. Both psychological stress and physiological stress can lead to cognitive impairment which negatively affects the student's academic performance, e.g., [38], [39], [40], [41][42]. The adoption of Metaverse services will allow the use of AI techniques that adjust the content, in real-time, to the current level of comprehension of the neurodivergent student and to her/his individual learning preferences. For example, adaptive learning algorithms can monitor and analyze the student's performance and level of engagement in real time and adjust the material presented to the student accordingly.

The major AI algorithms used to enable personalized learning environments include: (i) adaptive learning algorithms (used in DreamBox Learning[1] and Duolingo[2]), (ii) intelligent tutoring systems (used in Khanmigo[3] and Carnegie Learning[4]), and (iii) content recommendation algorithms (used in Coursera[5] and edX[6]).

**Sensory-Adjustable Learning Environments.**

Sensory-adjustable learning environments in the metaverse can be significantly enhanced through the integration of Brain-Computer Interfaces (BCIs) with VR headsets [43]. BCIs can monitor neural activity in real time to detect cognitive states such as attention, mental fatigue, or cognitive overload [44]. When combined with immersive VR learning environments, this capability allows the system to respond dynamically to a student's mental state [45]. For example, if the BCI detects signs of cognitive overload, such as sustained low attention levels or increased mental effort, the system can automatically reduce the difficulty of the learning content, simplify the visual environment, or suggest a short break involving calming or gamified activities. This adaptive regulation of cognitive load is particularly beneficial for neurodivergent students who may experience heightened sensitivity to sensory input or fluctuating attention [46]. By tailoring the learning experience based on real-time brain signals, BCI-enhanced VR systems offer a more personalized, responsive, and supportive educational environment that promotes sustained engagement and reduces stress and frustration

## 4.2 Improved Quality of Experience

To ensure metaverse-based education is effective and inclusive, it is essential to enhance the overall quality of the learner's experience. This involves creating environments that are not only engaging but also adaptable to individual needs. For neurodivergent students in particular, tailored features can significantly impact their ability to concentrate, engage, and thrive in immersive learning spaces.

**Personalized Sensory Settings.**

Personalized sensory settings in metaverse learning environments are essential for supporting neurodivergent students, who often experience heightened sensitivity to visual, auditory, or tactile stimuli. Using VR and AR technologies, learners can customize their virtual surroundings to minimize potential triggers of anxiety or sensory overload [47]. For example, students with autism or ADHD can adjust lighting, color contrast, ambient noise, and the number of environmental objects including decorative elements to create a calmer, more focused space [48].

---

[1] https://www.dreambox.com
[2] https://www.duolingo.com
[3] https://www.khanmigo.ai
[4] https://www.carnegielearning.com
[5] https://www.coursera.org
[6] https://www.edx.org

Spatial audio can be fine-tuned to reduce auditory distractions, while visual clutter can be minimized to help maintain attention. These settings can be configured prior to sessions and dynamically adapted in real time through AI-based monitoring of user behavior or via integrated physiological sensors [49]. By giving students control over their sensory input, metaverse platforms empower them to engage with educational content in ways that suit their individual comfort levels, this reducing stress, enhancing concentration, and fostering a more inclusive and effective learning experience.

**AI-Driven Engagement Tracking and Feedback Systems.**

In the context of education, traditional engagement tracking tools integrated in LMSs (e.g., Google Classroom[7], Canvas LMS[8], Blackboard Learn[9], and Schoology Learning[10]) are used to monitor and analyze how students interact with educational content and platforms delivering that content. For example, they may provide functionalities to track the engagement of students with course materials and activities, tools to conduct polls, and statistics on access patterns for each student. Usually, these engagement tracking tools are designed for neurotypical students. Neurodivergent students often have difficulties maintaining a proper level of focus and engagement when learning, particularly when using online platforms. They need frequent guidance while they learn. In the case of online delivery, neurodivergent students may face even greater challenges reducing their level of engagement such as sensory overload (e.g., complex interfaces and flashing screens [50]). New engagement tracking systems must therefore be developed to cater for the specific needs of neurodivergent students. An AI-driven engagement system can provide real-time personalized feedback that helps students correct any misunderstandings.

### 4.3   Enhanced Socialization and Collaboration

**Safe Virtual Spaces.**

Ensuring safe virtual spaces within the metaverse is crucial for fostering positive peer interactions and social learning, particularly for neurodivergent students who may struggle with traditional social environments. Controlled virtual settings provide structured, moderated spaces where students can engage in collaborative activities without the risks often associated with online interactions. These environments allow educators to implement safeguards such as identity verification, content moderation, and restricted access, ensuring that students interact in a secure and inclusive manner [22], [51].

---

[7] https://classroom.google.com/
[8] https://canvas.instructure.com
[9] https://help.blackboard.com/
[10] https://www.powerschool.com/solutions/personalized-learning/schoology-learning/

Beyond security measures, the design of these virtual spaces should prioritize accessibility and comfort, accommodating diverse communication preferences and social engagement styles. For instance, customizable avatars and text-based chat options can help students who experience anxiety in voice interactions. Additionally, structured social activities such as guided discussions, group projects, and role-playing exercises can facilitate meaningful peer engagement while reducing the stress of unstructured socialization [22], [52]

Safe virtual spaces also encourage the development of essential social skills in a risk-free environment. By engaging in simulated real-world interactions, neurodivergent students can practice collaboration, conflict resolution, and self-advocacy in a setting where mistakes do not carry real-world consequences. The ability to experiment with social behaviors in a controlled, supportive atmosphere fosters confidence and helps bridge the gap between virtual interactions and real-world socialization. As a result, well-designed virtual environments not only promote academic learning but also nurture social growth, empowering neurodivergent students to build stronger interpersonal connections in both digital and physical spaces [53], [54].

**Avatar-Based Communication Options.**

Avatar-based communication in metaverse learning environments offers a powerful solution for reducing social anxiety and enhancing engagement among neurodivergent students. By interacting through customizable avatars, learners can maintain a comfortable level of anonymity and control over their social presence, which helps alleviate the pressure of face-to-face communication [55]. This is particularly beneficial for students with autism or social anxiety, who may find traditional classroom interactions overwhelming. Avatars allow users to express themselves visually, choose how and when to engage, and participate in discussions without fear of judgment or misinterpretation [56]. Additionally, the use of non-verbal cues, spatial positioning, and visual expressions in avatar-based environments fosters more natural and inclusive interactions. This flexibility creates a safer and more supportive space for collaboration, peer learning, and community building which are key factors in boosting confidence, motivation, and academic success for neurodivergent learners.

### 4.4 Scalable and Cost-Effective Solutions

One of the key challenges in metaverse-based education is ensuring scalability and cost effectiveness, particularly for institutions with limited financial and technological resources. Currently, most schools cannot afford the acquisition costs of metaverse technologies (e.g., VR/AR) that can be used on a large scale (tens or hundreds of students). Future metaverse solutions for educating neurodivergent students must be developed with affordability as a prime requirement. Two key elements will help reduce costs: cloud-based deployments and the use of open-source metaverse frameworks.

**Cloud-Based Platforms.**

Typically, deploying a metaverse solution can cost $1 million or more depending on specific functionalities. Most K-12 schools and many colleges and universities may not afford the cost of acquiring and deploying metaverse devices to serve all their neurodivergent population. An alternative is to use a cloud-based metaverse which allows users to create and host immersive and persistent virtual worlds [57], [58]. A key benefit in using a public cloud provider for the metaverse is its scalability [58], which is a critical requirement for educational applications (accommodating a large number of students). Cloud platforms are also needed to provide the high-performance computing needed to enable metaverse applications. Moreover, the metaverse requires short response times to ensure that users enjoy a continuously seamless experience. This is particularly critical in the case of metaverse applications destined for neurodivergent students. Achieving a good response time is usually difficult with a small scale, on-premises deployment. A cloud deployment of metaverse educational platforms guarantees that neurodivergent students receive an effective learning experience.

**Open-Source Metaverse Frameworks.**

Open-source metaverse frameworks offer a sustainable solution by providing an accessible and flexible foundation for developing immersive learning environments without the high costs associated with proprietary platforms. These frameworks support widespread adoption in educational settings by allowing institutions to create tailored experiences that align with the diverse needs of neurodivergent students.

A crucial advantage of open-source metaverse frameworks is their ability to foster customization and flexibility. Educational institutions can modify these platforms to accommodate specific learning requirements, such as sensory-friendly interfaces, alternative interaction methods, and personalized avatars. Additionally, the collaborative nature of open-source development ensures continuous improvements, as global communities of developers contribute enhancements, security updates, and new features. This community-driven approach helps maintain system stability while addressing evolving educational needs.

Interoperability is another key benefit of open-source metaverse frameworks. Many of these platforms are designed to integrate seamlessly with existing LMSs, assistive technologies, and other digital education tools. By ensuring compatibility with a wide range of educational resources, open-source solutions promote smoother adoption and sustained engagement. Furthermore, open-source metaverse frameworks significantly reduce licensing costs, making them a more viable option for schools and universities that seek to implement metaverse-based learning environments without ongoing financial burdens.

Several open-source metaverse platforms have demonstrated their effectiveness in educational settings. For instance, OpenSimulator [59] provides a robust and flexible platform for creating customized virtual learning environments, enabling educators to design spaces tailored to neurodivergent students. Similarly, Vircadia supports decentralized virtual worlds, ensuring greater control over data privacy and user interactions [60]. By leveraging these scalable and cost-effective solutions, educational institutions can expand access to metaverse-based learning while maintaining affordability and adaptability.

## 5     Challenges and Risks

While metaverse services hold great promise for enhancing immersive and interactive learning experiences, their widespread adoption in education comes with several challenges and risks. Ensuring the privacy and security of student data is paramount, particularly for neurodivergent learners who may require personalized support mechanisms. Additionally, disparities in access to technology raise concerns about digital equity and accessibility, potentially exacerbating educational inequalities. The cognitive load associated with complex virtual environments must also be carefully managed to prevent overwhelming students. Furthermore, the scalability and interoperability of metaverse services remain critical for seamless integration with existing educational technologies. Finally, effective educator training and implementation strategies are necessary to ensure that teachers can leverage these tools to their full potential. This section explores these key challenges and their implications for metaverse-based learning.

### 5.1    Privacy and Data Security

Ensuring privacy and data security in metaverse-based learning is particularly critical for neurodivergent students, who often belong to vulnerable populations that require additional safeguards. These students may have specialized learning profiles incorporating detailed behavioral, cognitive, and personal information, making the risk of data exposure a significant concern. Without robust security measures, sensitive data becomes vulnerable to cyberattacks or unauthorized access, potentially leading to stigma, discrimination, or exploitation. Research highlights the need for comprehensive data governance and tailored security protocols that address the unique vulnerabilities of neurodivergent learners [61].

To mitigate these risks, there is a growing emphasis on decentralized authentication mechanisms and secure storage solutions. Traditional centralized systems typically concentrate data in one location, presenting a single point of failure which is a particularly dangerous scenario when managing the sensitive information of neurodivergent students. In contrast, decentralized approaches, such as blockchain-based identity management and federated learning, distribute data control across multiple nodes. This strategy minimizes the likelihood of unauthorized access and ensures that any breach is localized, thereby protecting the integrity and confidentiality of each student's data [62].

Moreover, implementing decentralized authentication and secure storage solutions is essential for ensuring compliance with global data protection regulations such as GDPR and ERPA. These regulations mandate rigorous controls over how personal data is collected, processed, and stored, a requirement that is particularly crucial when dealing with the sensitive learning profiles of neurodivergent students. By leveraging decentralized security protocols, educational institutions can create a secure and inclusive metaverse environment that supports adaptive learning while building trust among students, educators, policymakers, and other stakeholders.

While federated learning and blockchain strengthen privacy protections, they can also introduce computational overhead and latency challenges. To balance privacy preservation with efficiency, hybrid approaches are increasingly being explored. For example, combining federated learning with edge computing allows sensitive computations to occur locally while offloading intensive tasks to the cloud, reducing lag in VR/AR environments. Similarly, lightweight blockchain protocols or sidechains can be employed for identity verification without burdening real-time interactions. These strategies ensure that privacy safeguards remain robust while maintaining the low-latency, responsive performance that immersive educational applications require, particularly for neurodivergent learners who may be more sensitive to delays or disruptions.

### 5.2 Digital Equity and Accessibility

Another challenge when providing metaverse educational services to neurodivergent students is to ensure digital equity, that is, ensuring that all students have the skills, technology, and Internet service so that they may fully benefit from the metaverse. To enable equitable access, metaverse educational services must adhere to accessibility standards such as Web Content Accessibility Guidelines (WCAG) 2.2 [37]. This allows users with cognitive impairment to access and properly use metaverse services. Also, students must have access to specialized technologies (e.g., VR/AR headsets) needed for the metaverse. To ensure equitable access, formal frameworks must be adopted. An example is the Universal Design for Learning (UDL) which define guidelines to be used to "improve and optimize teaching and learning for all people" [63].

### 5.3 Cognitive Load Management

Cognitive load refers to the amount of mental effort required to process information and complete tasks [64]. Cognitive Load Theory (CLT) identifies three types of cognitive load [65]: (i) intrinsic (inherent difficulty of the material), (ii) extraneous (unnecessary mental effort caused by an inadequate presentation of the information), and (iii) germane (mental effort needed to understand and store the information). Research shows that neurodivergent learners experience a higher extraneous cognitive load due to sensory sensitiveness (as the brain focuses on sensory processing, this may shut off other functions such as information processing). As a result, neurodivergent students do not react to immersive environments as their neurotypical peers do. They are usually sensory sensitive, i.e., they have a heightened awareness and reactivity to sensory stimuli [66]. As a result, these students are more susceptible to sensory overload in immersive environments, which occurs when the brain becomes overwhelmed by excessive stimulation. For example, a 2024 experimental study found that autistic students reported experiencing higher sensory issues, leading to poor academic outcomes [67].

To reduce cognitive load, metaverse services for neurodivergent students must include adaptive filtering mechanisms. These mechanisms must provide a wide range of features, such as visual/auditory filtering (adjusting visual complexity, brightness, contrast, volume, and selection of audio sources), contextual filtering (focusing on the current learning objective), breaking down information into manageable chunks, and adjusting the pace and format of information presentation.

Adaptive filtering techniques include: (i) brain-computer interface (BCI) integration, (ii) real-time sensory adjustment, and (iii) AI-driven adaptive learning. BCIs continuously monitor a person's brain activity and adjusts the learning tasks according to the current mental effort of the student. Real-time sensory adjustment uses sensors to monitor a student's physiological state and triggers changes (e.g., reduce the volume) or alerts (e.g., to the educators) whenever it detects signs of stress, overstimulation, anxiety, or disengagement. AI-driven adaptive learning systems (such as Lexia[11] and Nearpod[12]) analyze data on student performance (e.g., accuracy of answers and response time) and select the most appropriate learning content for that student. The content may be drawn from an existing library or generated in real time.

Recent research suggests that these adaptive filtering mechanisms can be enhanced through the integration of advanced tools. For example, BCIs integrated with VR headsets can monitor neural signals in real time to detect indicators of cognitive overload, such as reduced attention or increased mental effort, and automatically adjust the complexity or pace of the learning environment [43], [44], [46]. Similarly, real-time sensory adjustment systems embedded in VR/AR APIs allow on-the-fly modification of brightness, contrast, and auditory stimuli, reducing the risk of sensory overload. Some platforms also incorporate physiological sensors (e.g., eye-tracking and heart rate variability) to dynamically regulate information density and provide calming interventions when stress is detected. Together, these tools demonstrate how adaptive filtering can move beyond static adjustments to intelligent, real-time regulation of cognitive load, thereby creating more sustainable and supportive learning experiences for neurodivergent students.

### 5.4 Service Scalability and Interoperability

For metaverse-based learning to be adopted at scale, it must seamlessly integrate with existing educational technologies such as LMSs, assessment tools, and digital content repositories. This requires the development of open APIs, standardized data formats, and interoperability protocols that enable smooth communication between immersive platforms and conventional systems [68]. Moreover, to ensure equitable access for neurodivergent learners, universal accessibility standards must be established across devices and platforms. These standards should address not only physical accessibility but also cognitive and sensory needs, enabling personalized learning experiences to remain consistent regardless of the technical ecosystem. Importantly, metaverse platforms should support both Android-based and Windows-based headsets, including low-cost, standalone Android VR devices that do not require connection to a high-performance computer. This flexibility ensures broader access, especially in low-resource educational settings, and allows students and institutions to adopt hardware solutions that best fit their budget and infrastructure. Achieving such interoperability and scalability is critical to promoting widespread, cost-effective deployment of inclusive metaverse learning environments.

---

[11] https://www.lexialearning.com
[12] https://nearpod.com

### 5.5 Educator Training and Implementation Challenges

A significant barrier to the effective integration of metaverse services in education is the current lack of educator readiness and expertise in immersive technologies. Many instructors are unfamiliar with designing, navigating, or facilitating learning within 3D virtual environments, particularly when tailoring experiences for neurodivergent students [69]. To overcome this, targeted professional development programs are essential. These programs should cover both the technical skills required to operate metaverse platforms and the pedagogical strategies for fostering inclusive, student-centered learning in virtual spaces. Providing educators with hands-on experience, instructional design frameworks, and accessibility training will be key to ensuring the successful and sustainable adoption of metaverse-based education. One promising solution is the use of generative AI, which can assist educators in automatically generating, expanding, and customizing virtual learning environments without requiring programming or 3D modeling expertise [70]. With intuitive prompts and AI-driven design tools, teachers can create immersive lessons and environments tailored to their course objectives and student needs, significantly lowering the barrier to entry for non-technical users.

To further strengthen educator readiness, professional development should combine both technical and pedagogical training. On the technical side, workshops and certification programs in VR/AR platforms (e.g., Unity Learn, Unreal Engine Education, and Google Expeditions training modules) can provide hands-on experience in designing and navigating immersive environments. On the pedagogical side, frameworks such as Universal Design for Learning (UDL) and differentiated instruction should be integrated into training so teachers can adapt immersive lessons to diverse cognitive and sensory profiles. Additional strategies include co-design workshops where educators collaborate with neurodivergent students to build accessible lesson plans, as well as micro-credentialing programs that recognize mastery of inclusive immersive teaching practices. Blended models, combining online tutorials with live mentoring, have also been shown to improve teachers' confidence in applying immersive technologies in real classrooms. Together, these targeted approaches can equip educators to implement the proposed framework in ways that are technically competent, pedagogically inclusive, and responsive to neurodivergent learners' needs.

# 6  Recommendations and Policy Considerations

To fully harness the potential of metaverse services in education while addressing the associated challenges, a strategic approach is needed to ensure security, inclusivity, and long-term sustainability. Privacy-first metaverse service design must be prioritized to protect sensitive student data, particularly for neurodivergent learners who rely on personalized support. Additionally, the development of inclusive learning standards can help create accessible and adaptable virtual environments that cater to diverse cognitive and sensory needs. Ensuring interoperability and the adoption of open standards will be crucial for seamless integration with existing educational technologies and service-oriented systems. Finally, institutional and government support is essential in fostering policies, funding, and training programs that enable educators and learners to effectively engage with metaverse-based education. This section outlines key recommendations and policy considerations for creating a secure, inclusive, and scalable metaverse learning ecosystem for neurodivergent students.

## 6.1  Privacy-First Metaverse Service Design

One of the most critical concerns in metaverse-based education is integrating privacy considerations into the development of metaverse services from their inception. This proactive approach helps identify and mitigate privacy risks early in the design process. Hamadi et al. [71] highlighted the necessity of incorporating privacy considerations, such as data collection, disclosure, and retention, into service design.

Another critical concern in metaverse-based education is ensuring that AI-driven personalization does not compromise student privacy. Traditional machine learning models require centralized data storage, which increases the risk of exposing sensitive information, particularly for neurodivergent students who may have unique learning profiles and accessibility needs. Federated learning offers a privacy-preserving alternative by allowing AI models to be trained locally on student devices or within institutional networks without transmitting raw data to external servers [61]. This decentralized approach ensures that personalization features, such as adaptive content recommendations, real-time feedback mechanisms, and cognitive load adjustments, can be implemented while minimizing the risk of unauthorized access or data breaches.

For neurodivergent students, federated learning enables tailored learning experiences without requiring them to surrender personal data to centralized databases, which could otherwise expose them to risks of profiling or discrimination. AI models trained using federated learning can recognize patterns in user engagement, adjust lesson delivery formats, and refine sensory-friendly settings without retaining identifiable user information [72]. By embedding these privacy-aware AI methods into metaverse services, educators and developers can enhance student engagement while maintaining strict data protection standards. Moreover, federated learning aligns with ethical guidelines for educational technology, ensuring that personalization benefits all learners without compromising their rights to data security and autonomy.

In addition to implementing privacy-preserving AI methods, establishing student-centered data privacy policies is essential for fostering trust and ensuring compliance with global data protection regulations, such as GDPR and FERPA [73]. For neurodivergent students, privacy concerns extend beyond conventional data security measures. They may require additional protections related to behavioral data, assistive technology usage, and communication preferences. Student-centered policies should prioritize transparency, giving learners and their guardians clear insights into how their data is collected, used, and stored.

A core principle of student-centered privacy policies is informed consent and data ownership. Neurodivergent students and their families should have control over what data is shared, with the ability to opt in or out of specific data-driven services. Additionally, data minimization techniques, such as only storing essential learning analytics rather than full behavioral logs, can help reduce exposure risks. Implementing role-based access control (RBAC) ensures that only authorized educators and support staff can access sensitive information, preventing unnecessary data visibility and potential misuse [74].

Furthermore, privacy policies should be adaptive to individual learning needs, meaning they must evolve alongside advancements in AI and metaverse technologies. Continuous collaboration with neurodivergent individuals, educators, policymakers, and privacy experts can ensure that data policies remain ethical, effective, and aligned with the latest accessibility standards. By adopting student-centered privacy frameworks, metaverse-based learning platforms can promote a secure, equitable, and empowering digital learning experience, where neurodivergent students can engage confidently without concerns about privacy infringement.

### 6.2 Implementation Pathways

While our framework has been presented conceptually, it is also designed with practical implementation in mind. For personalized learning environments, reinforcement learning algorithms (e.g., Q-learning or Deep Q-Networks) can dynamically adapt the sequencing and difficulty of activities, while collaborative filtering methods and transformer-based NLP tutors recommend resources and provide conversational support aligned with learner preferences. Multimodal interaction can be enabled through AI-driven gesture recognition models for VR/AR input, convolutional neural networks for motion tracking, and adaptive speech recognition to accommodate diverse communication styles.

To support service modelling and delivery, orchestration platforms such as Kubernetes with edge extensions ensure low-latency VR/AR rendering, while game engines (e.g., Unity or Unreal) integrate reinforcement learning APIs to dynamically reconfigure content. For privacy-preserving infrastructure, federated learning protocols (e.g., FedAvg, FedProx) and differential privacy mechanisms (Laplace or Gaussian noise injection) can safeguard student data while enabling personalization, complemented by blockchain-based smart contracts for decentralized identity management. Finally, cross-domain interoperability can be realized through open standards like Learning Tools Interoperability (LTI), middleware synchronization of learning progress across devices, and blockchain-backed credentialing for secure transferability of achievements across platforms and institutions.

Together, these implementation pathways demonstrate that the framework is not only theoretically robust but also practically operationalizable, bridging inclusive design principles with deployable AI, privacy, and interoperability technologies.

## 6.3 Evaluation and Assessment Considerations

To assess the effectiveness of the proposed framework, a multi-dimensional evaluation strategy is required that combines learning outcomes, engagement indicators, and inclusivity measures specifically tailored for neurodivergent students.

Learning Outcomes: Academic performance can be measured through pre- and post-test score improvements, accuracy in task completion, and knowledge retention. To ensure that immersive environments do not increase cognitive strain, standardized instruments such as the NASA-TLX and the Paas Cognitive Load Rating Scale can be applied. Adaptive assessments embedded within the metaverse environment further allow ongoing monitoring of skill acquisition and progression.

Engagement: Engagement should be assessed across behavioral, physiological, and self-reported dimensions. Behavioral metrics include frequency and duration of participation and interaction with multimodal tools. Physiological and affective engagement can be monitored through biosensors or Brain-Computer Interfaces (BCIs), which capture indicators such as attention levels, stress, or cognitive effort. Complementing these measures, validated self-report instruments such as the User Engagement Scale (UES) can provide insights into learners' subjective experience.

Inclusivity and Accessibility: Inclusivity can be evaluated by aligning the learning environment with Universal Design for Learning (UDL) guidelines and testing accessibility against WCAG 2.2 standards. Social inclusion and comfort with interaction modes can be captured through instruments such as the Classroom Community Scale, while qualitative insights can be obtained from interviews and focus groups with students, educators, and guardians.

Furthermore, these metrics can be directly mapped to the five framework components (see Section 3), ensuring that personalization, multimodal interaction, secure infrastructure, service delivery, and interoperability are each evaluated with context-appropriate measures.

### 6.4 Development of Inclusive Metaverse Learning Standards

Several organizations are currently developing standards for the metaverse. They are focusing on different aspects relevant to the metaverse, including interoperability, identity and privacy, and security. For example, the IEEE has developed several relevant standards, including IEEE P2048 (terminology, definitions, and taxonomy), IEEE 3812.1 (general requirements for identity framework for metaverse), and IEEE P7016 (ethically aligned design and operation of metaverse systems). The ISO/IEC joint technical committee 1 (JTC 1) is working on standards related to technologies vital for creating immersive metaverse experiences, for instance, virtual world representation and human animation [75]. Most of the standardization efforts have hitherto focused on interoperability. For example, the Metaverse Standards Forum, which started in April 2023, is a consortium of over 2400 standards organizations (including W3C, IEEE, and Khronos Group) and major companies (e.g., Meta, Microsoft, NVIDIA, Sony, Adobe, and Huawei). The major focus of this forum is to foster interoperability standards for an open metaverse [76].

While current standardization initiatives will contribute to building future robust metaverse systems, a critical aspect where standards are missing is inclusion. Future metaverse learning environments must be built following learning standards that satisfy the specific needs of neurodivergent students. Metaverse inclusion standards must ensure accessibility (e.g., that interfaces are readily usable by neurodivergent learners with visual, audio, or cognitive impairments). Recognizing the importance of inclusive metaverse learning systems, several organizations have started efforts to develop accessibility and inclusion guidelines in metaverse environments. For example, the International Telecommunications Union (ITU) is developing a set of guidelines and recommendations to integrate accessibility products and services in the metaverse for people with diverse access needs [77].

### 6.5 Interoperability and Open Standards

To ensure that metaverse-based educational tools can be widely adopted and effectively integrated into existing digital ecosystems, interoperability and adherence to open standards are essential. The adoption of open APIs enables seamless integration with widely used educational platforms, including LMSs, content repositories, and communication tools [78]. This allows educators and institutions to incorporate immersive experiences without overhauling their existing infrastructures. Additionally, standardized cross-platform learning environments ensure that educational content and interactions are consistent across a wide range of devices and operating systems. Such standards are particularly important for maintaining accessibility and usability for neurodivergent students, whose learning experience must remain stable and personalized across different hardware configurations. Open standards also encourage innovation by allowing third-party developers to build compatible tools and extensions, ultimately fostering a more flexible, inclusive, and sustainable metaverse learning ecosystem.

### 6.6 Institutional and Government Support

Building metaverse learning platforms that provide adequate services to neurodivergent learners requires substantial investments from public and private institutions. In the US, this is often achieved through grant programs funded by the federal and/or state governments. The Digital Equity Act authorized $2.75 billion in funding for programs designed to promote digital equity and inclusion [79]. The goal of this act is to prevent digital exclusion, i.e., scenarios where advances such as the metaverse become an exclusive space accessible only to segments of the population that can afford metaverse technologies. In Australia, the government provides funding through the Higher Education Disability Support Program to assist students with disabilities (including neurodivergent students) in higher education [80].

The broad adoption of metaverse learning services will require substantial investment in training teachers in the area of immersive learning. A 2024 study estimates that over 40% of K-12 schools in the US incorporated AR/VR technologies in 2024, from less than 20% in 2022 [81]. Despite this remarkable market penetration, teachers are still not adequately trained to benefit from immersive technologies. A 2022 large scale (N = 20876) study revealed that the lack of adequate training and resources are significant barriers to widespread VR/AR use in K-12 classrooms [82]. Teachers need to know how to integrate immersive technologies into their curriculum and how to use these technologies to effectively benefit neurodivergent students.

To translate these recommendations into practice, policymakers and institutions can adopt several actionable measures. First, funding models should include multi-tiered support: short-term seed grants for pilot projects, medium-term institutional investment in infrastructure and training, and long-term government or philanthropic funding streams to ensure sustainability. Second, a phased timeline can be implemented, beginning with small-scale pilots in selected schools or universities, followed by regional rollouts once feasibility and inclusivity outcomes are validated, and culminating in national adoption supported by standardized guidelines. Third, strategic partnerships with technology providers, educational consortia, and disability advocacy organizations can accelerate development while ensuring that diverse learner needs are represented in design and evaluation. Finally, cross-sector collaboration between governments, universities, and industry can create shared infrastructure and training hubs, reducing duplication and ensuring equitable access to resources. Together, these steps provide a pragmatic roadmap for turning high-level principles into implementable policies that support the widespread and inclusive adoption of metaverse-based education.

### 6.7 Equity and Adaptation for Low-Resource Settings

While Subsection 4.4 discussed scalability and cost-effective solutions, an equally important consideration is ensuring that the proposed framework can be equitably implemented in low-resource contexts where access to advanced VR/AR technologies may be limited. Without such adaptations, there is a risk that the framework could primarily benefit well-resourced institutions while leaving behind students in underserved regions.

To address this, the framework can be adapted through several strategies. First, tiered deployment models can allow immersive content to be accessed through lightweight devices such as smartphones, tablets, or low-cost standalone VR headsets, supported by WebXR platforms that run in standard browsers without specialized hardware. Second, cloud-based rendering and streaming can offload processing to remote servers, reducing local device requirements, while downloadable or asynchronous modules can provide continuity in bandwidth-constrained environments. Third, institutions can leverage open-source platforms (e.g., OpenSimulator, Vircadia) that reduce licensing costs and enable customization to local pedagogical and cultural needs. Partnerships with governments, NGOs, and industry can further expand access through community VR labs, device lending programs, or subsidized connectivity initiatives.

Finally, the framework emphasizes inclusivity, personalization, and multimodality, principles that are not inherently dependent on high-end VR/AR systems. Even in low-resource settings, adaptive AI-driven personalization, accessible multimedia content (text, audio, video), and compliance with UDL and WCAG guidelines can be implemented to ensure equitable opportunities for neurodivergent learners.

Together, these adaptations ensure that the framework is not only scalable but also sensitive to global disparities in access, enabling immersive and inclusive learning experiences across diverse institutional and regional contexts.

## 7 Conclusion

In this paper, we discussed the transition from service-oriented computing to metaverse services which represents a significant evolution in digital education, particularly for neurodivergent students. Traditional service-oriented computing platforms, while modular and scalable, often lack the immersive and adaptive qualities necessary for personalized learning. Metaverse-based environments, by contrast, offer real-time, AI-driven personalization, multi-sensory engagement, and persistent, context-aware learning experiences. These advancements enable a more flexible and inclusive approach to education, accommodating diverse cognitive needs through immersive virtual spaces, adaptive feedback mechanisms, and cross-platform integration.

Metaverse services hold significant promises for enhancing the educational experiences of neurodivergent students. By providing customizable learning environments, facilitating active learning strategies, and fostering peer support, they can address many of the barriers these students face in traditional educational settings. However, to fully realize the benefits of metaverse learning and successful implementation, challenges related to privacy, accessibility, and commitment to inclusivity must be addressed through continued research, policy development, educators training, and

4interdisciplinary collaboration to ensure that all students can thrive in these innovative learning spaces.

We also proposed a generic framework that illustrates how metaverse services can transform and improve education for neurodivergent students by providing personalized, multimodal, and privacy-aware learning environments. By leveraging AI-driven personalization, immersive VR/AR experiences, adaptive content delivery, and robust privacy safeguards, this framework ensures that neurodivergent learners can thrive in engaging, distraction-free, and flexible educational settings.

To advance the adoption of metaverse-based learning for neurodivergent students, we also believe a multi-stakeholder approach is necessary. Educators, policymakers, and technologists must collaborate to establish accessibility standards, ensuring that immersive learning environments are inclusive and adaptable to diverse needs. Further research is required to refine AI-driven personalization, mitigate cognitive overload, and evaluate long-term learning outcomes in metaverse spaces. Additionally, educational institutions should invest in professional development programs to equip teachers with the skills needed to implement and manage virtual learning environments effectively. Finally, privacy and security frameworks must be reinforced to protect student data while allowing for adaptive and responsive learning experiences.

## References

[1]    HealthBeat, "Understanding Neurodiversity: Exploring Differences in Brain Function," 2023.
[2]    Australian Institute of Health and Welfare, "People with Disability in Australia," 2024.
[3]    G. Cadby, T. Pitman, and P. Koshy, "Students with Disability in Australian Higher Education (Nov 2024 Update)," 2024.
[4]    CDC, "Data and Statistics on ADHD. Centers for Disease Control and Prevention (CDC)," 2024.
[5]    Yale, "The Yale Center for Dyslexia and Creativity. [Online] available: https://dyslexia.yale.edu," 2025.
[6]    S. Fletcher-Watson and F. Happé, *Autism: A new introduction to psychological theory and current debate*. Routledge, 2019.
[7]    M. P. Papazoglou, P. Traverso, S. Dustdar, and F. Leymann, "Service-oriented computing: a research roadmap," *Int J Coop Inf Syst*, vol. 17, no. 02, pp. 223–255, 2008.
[8]    C. Dede, "Immersive interfaces for engagement and learning," *Science (1979)*, vol. 323, no. 5910, pp. 66–69, 2009.
[9]    X. Xu *et al.*, "Metaverse services: The way of services towards the future," in *2023 IEEE International Conference on Web Services (ICWS)*, IEEE, 2023, pp. 179–185.
[10]   L.-H. Lee *et al.*, "All one needs to know about metaverse: A complete survey on technological singularity, virtual ecosystem, and research agenda," *Foun-*

2828